# Visualizing Nanoscopic Acoustic Mode Competition in van der Waals Ferroelectric

Zhaodong Chu[1], Carter Fox[2], Zixin Zhai[3], Haihua Liu[4], Priti Yadav[3], Bing Lv[3], Yue Li[1], Thomas E Gage[4,*], Jun Xiao[2,*], Haidan Wen[1,5,*]

[1]Materials Science Division, Argonne National Laboratory, Lemont, IL 60439, USA

[2]Department of Materials Science and Engineering, University of Wisconsin–Madison, Madison, WI, USA

[3]Department of Physics, The University of Texas at Dallas, Richardson, TX 75080, USA

[4]Center for Nanoscale Materials, Argonne National Laboratory, Lemont, IL 60439, USA

[5]Advanced Photon Source, Argonne National Laboratory, Lemont, IL 60439, USA

*emails: tgage@anl.gov; jun.xiao@wisc.edu; wen@anl.gov

**Abstract**

Understanding how low-dimensional ferroelectrics respond to ultrafast excitation at nanoscales is essential for controlling energy flow and mechanical functionality in next-generation polar devices, yet the nanoscopic structural response to ultrafast depolarization remains unresolved, obscuring the microscopic pathways of acoustic decoherence and energy dissipation. Here, we spatiotemporally resolve lattice motion in the van der Waals ferroelectric $NbOI_2$ using combined ultrafast electron microscopy and diffraction, revealing three acoustic phonons: two transverse shear modes and one longitudinal breathing mode. The transverse mode that shears the layers perpendicular to the in-plane polar axis dominates over that along the polar axis, reflecting anisotropic polarization-strain coupling. Real-space mapping uncovers spatially correlated heterogeneity in mode amplitudes and lifetimes. Regions dominated by a single shear mode exhibit significantly longer acoustic lifetimes than multimode regions, suggesting acoustic phonon–phonon scattering as a major source of decoherence. Our results provide a microscopic understanding of ultrafast depolarization-driven acoustic dynamics and spatially heterogeneous energy dissipation in van der Waals ferroelectrics.

**MAIN TEXT**

**Introduction**

Van der Waals (vdW) ferroelectrics have recently emerged as a versatile platform for exploring low-dimensional polarization phenomena with opportunities in non-volatile memory, logic, sensing, and optoelectronics[1-5]. In contrast to conventional three-dimensional ferroelectrics, weak interlayer vdW coupling and the mechanical flexibility of heterostructure stacking allow robust polarization to persist down to the atomic-thin limit, either via exfoliation from bulk crystals[6-10] or through moiré engineering[11–13]. The coexistence of long-range dipolar order and low-symmetry lattice distortion gives rise to unusually strong electromechanical and nonlinear-optical responses[14-16], making vdW ferroelectrics an ideal testbed for studying how polarization, charge, and strain interact under nonequilibrium excitation. Understanding these interactions on ultrafast timescales is central to advancing piezo-optoelectronic and optoferroic functionalities in next-generation nanoscale devices[17-19].

The vdW ferroelectric niobium oxyiodide ($NbOI_2$)[20-22] has recently attracted significant attention as a platform hosting strong coupling among polarization, charge, lattice, and optical excitations. It crystallizes in a low-symmetry monoclinic C2 structure, in which an off-centering of Nb along the *b*-axis establishes an in-plane ferroelectric polarization ***P*** with a Curie temperature of ~ 189 °C [22]. Optical studies have revealed giant second-harmonic generation[23] and colossal terahertz (THz) emission[24, 25], reflecting its large second-order susceptibility and efficient light–polarization conversion. Additionally, the observation of ferrons[26] reveals ultrafast collective ferroelectric dynamics in $NbOI_2$. These advances suggest $NbOI_2$ as a unique vdW ferroelectric where electronic and structural degrees of freedom are intimately coupled on ultrafast timescales. Very recent ultrafast diffraction measurements[27] have shown that above-band-gap photoexcitation transiently suppresses the in-plane polarization and generates coherent acoustic strain waves in $NbOI_2$. These reciprocal-space measurements, averaging over large regions, are difficult to reveal the spatial heterogeneity of acoustic modes and their decay lifetimes. The real-space dynamical imaging capability of ultrafast electron nanoprobes has enabled visualization of carrier and lattice dynamics in non-polar semiconductors[28-30]. However, imaging heterogeneous dynamics in vdW polar systems, which is essential for understanding ultrafast polarization–strain coupling and energy flow in low-dimensional ferroelectrics on nanoscales, with implications for designing and operating high-speed electromechanical, optoferroic, and phononic nanodevices based on vdW materials, remains unexplored.

In this work, we combine ultrafast electron microscopy (UEM) and diffraction (UED) to obtain a spatiotemporal view of lattice dynamics in freestanding $NbOI_2$ flakes. Upon above-band-gap excitation, we observe three coherent acoustic modes: two transverse acoustic (TA) shear modes and one longitudinal acoustic (LA) breathing mode. Real-space UEM imaging directly visualizes nanoscale bend contour oscillations associated with these modes and reveals pronounced spatial variations in both amplitude and decay times. By analyzing the dynamics of the reciprocal lattice using UED, the shear displacement is found to be perpendicular to the in-plane polar axis and preferentially excited, reflecting the strongly anisotropic coupling between polarization suppression and shear strain in this low-symmetry ferroelectric. Moreover, multimode regions exhibit shorter decay times, indicating spatially heterogeneous elastic energy dissipation in $NbOI_2$. Ultrafast depolarization drives the inverse piezoelectric response that is mainly responsible to excite the two transverse shear modes, whereas the longitudinal breathing mode is driven by thermoelastic stress. Our results provide direct insight into the nanoscale heterogeneity of ultrafast elastic dynamics in $NbOI_2$, paving the way for engineered strain control in layered ferroelectrics.

**RESULTS**

**Visualization of nanoscale acoustic motion**

Our experiments combine UEM and UED to image the structural dynamics of freestanding $NbOI_2$ flakes following ultrafast excitation, as shown in the schematic in Fig.1 (a). The sample is excited by a 400 fs, linearly polarized laser pulse at a photo energy of 2.41 eV, above the $NbOI_2$ indirect bandgap (2.24 eV)[23], with a fluence of 3 mJ/cm$^2$. The lattice dynamics are probed using 200 keV electron pulses, providing ~ 1 ps temporal and ~ 1 nm real-space resolution. Optical, thickness, and Raman characterization of the $NbOI_2$ flake are provided in section S1 and fig. S1. As illustrated in the set of Fig.1(a), photo-generated hot carriers rapidly generate a transient screening field that suppresses the ferroelectric polarization, with lattice heating occurring subsequently. This ultrafast polarization suppression is directly evidenced in Fig. 1(b) by the observed transient beam shift of the central (000) non-diffracted spot toward the –***b**** direction at a pump-probe delay of 1 ps in a ~100 nm-thick freestanding $NbOI_2$ flake measured under the [201] zone axis. This amount of shift suggests a sudden reduction of the polarization ($\Delta \boldsymbol{P}$)-induced internal electric field $\Delta \boldsymbol{E}$ by approximately $7 \times 10^8$ V/m, and the suppression fully recovers within a few picoseconds (section S2 and fig. S2), consistent with the previous UED experiments[27].

Following ultrafast polarization suppression and lattice heating, the resulting stress generates coherent lattice dynamics that manifest as bend contour motion in UEM real-space images. Fig. 1(c) shows a bright-field UEM image of the 100 nm-thick freestanding $NbOI_2$ flake under the [201] zone axis prior to excitation. A video of the real-space contour oscillations following laser excitation is provided in section S3 (Movie S1). The laser polarization is aligned along the crystallographic c-axis to maximize optical excitation in all UEM and UED measurements presented in the main text, owing to the strongly anisotropic optical absorption of NbOI$_2$ (section S4 and fig. S3). Additionally, no contour oscillations or coherent acoustic responses are observed under below-band-gap excitation at 1.2 eV, confirming that photo-carrier generation is essential for launching the acoustic motion. To probe the local structural dynamics, we track the contour motion along four representative line cuts indicated by colored arrows in Fig. 1(c). The spatiotemporal intensity profiles along these line cuts (Fig. 1(d–g), left panels) display clear oscillatory modulations that decay over time. The corresponding Fast Fourier Transform (FFT) spectra (Fig. 1(d–g), right panels) reveal that, depending on position, the contour oscillation is composed of either a single frequency or multiple frequencies. For example, along the blue and red line cuts, the dynamics are dominated by modes $f_1 = 8.0 \pm 0.3$ GHz and $f_2 = 9.3 \pm 0.3$ GHz, respectively, whereas some regions along the other two-line cuts exhibit three modes, including a higher-frequency component at $f_3 = 12.7 \pm 0.3$ GHz. Here, the uncertainty of the mode frequency is set by the inverse of the finite time window of the measurement (1.5 ns). This spatial variation in the local frequency content shows that the projected elastic response captured by UEM varies strongly across the flake. This motivates a more detailed reciprocal-space analysis to identify the nature and symmetry of the underlying acoustic modes.

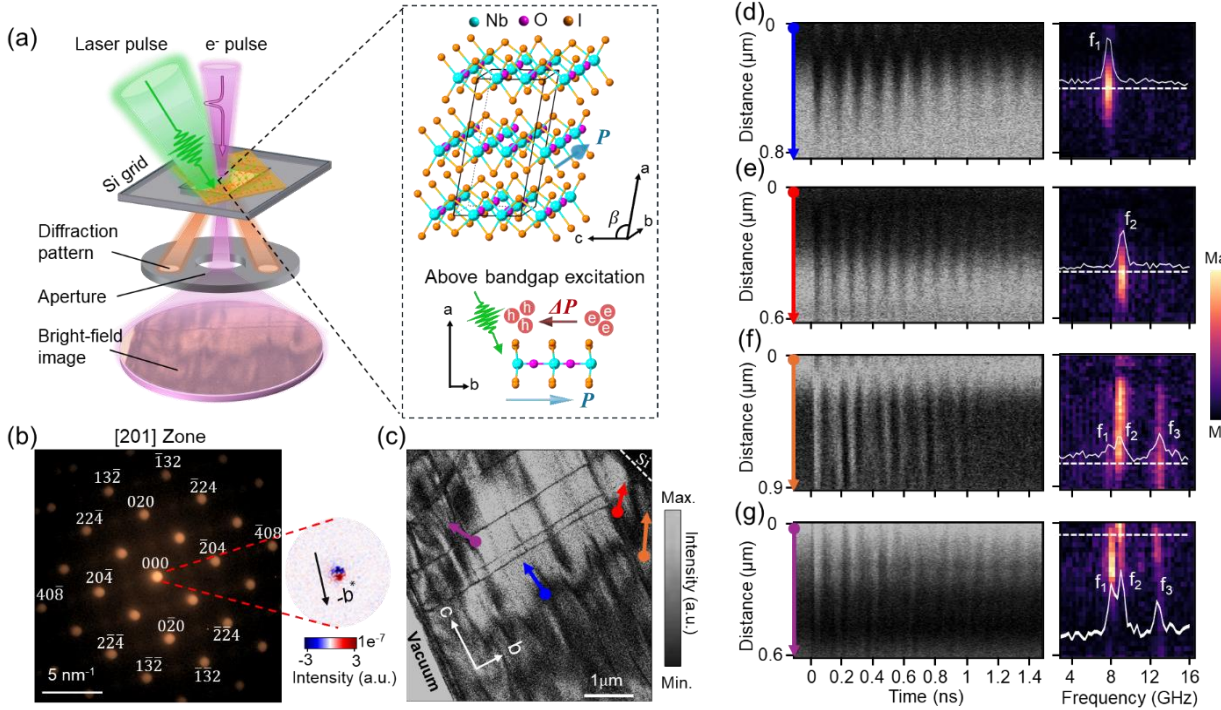

*Fig. 1. Ultrafast depolarization and real-space visualization of coherent acoustic waves in freestanding NbOI$_2$.* *(a) Schematic of the combined UED and UEM experiments on a freestanding NbOI$_2$ flake. The experiments are performed at room temperature. The dashed-box inset depicts the monoclinic crystal structure of NbOI$_2$ (a = 1.518 nm, b = 0.392 nm, c = 0.752 nm, β = 105.5°), and the in-plane polarization **P** along the b-axis, transiently suppressed by above-band-gap excitation. The*

*resulting lattice dynamics are probed by UED and UEM. (b,c) Static electron diffraction pattern and bright-field UEM image of the same ~100 nm-thick flake under the [201] zone axis. In (b), Bragg peaks are indexed, and the right inset (1.5× enlarged) shows the differential diffraction pattern of the central (000) beam at 1 ps relative to 0 ps, revealing a transient shift toward −**b**\* and evidencing ultrafast polarization suppression. In (c), colored arrows mark representative line cuts used to extract local real-space dynamics. The rotation of sample orientation between diffraction and real-space images arises from the UED/UEM setup configuration. (d–g), Spatiotemporal intensity profiles (left) and corresponding FFT spectra (right) along the line cuts, revealing position-dependent contour oscillations containing one or more acoustic modes. The line-cut FFT profiles along the dashed lines show the three dominant frequencies $f_1$, $f_2$, and $f_3$.*

**Reciprocal-space identification of three coherent acoustic modes**

To assign the symmetry and displacement character of the coherent acoustic modes, we analyze time-dependent diffraction intensities of selected Bragg reflections using UED. The electron beam probes the sample along the [201] zone axis, tilted by ~ 1.2° from the sample surface normal, providing sensitivity to both transverse and longitudinal acoustic motions. We focus on three representative Friedel Bragg peak pairs: $(\bar{4}08, 40\bar{8}; 0\bar{2}0, 020; \bar{1}3 2, 13\bar{2})$, because they provide complementary sensitivities to distinct components of the lattice response. The $(\bar{4}08, 40\bar{8})$ pair with $k = 0$ is primarily sensitive to distortions in the a-c plane, particularly β-angle change. The $(0\bar{2}0, 020)$ pair, with $h = l = 0$, probes dynamics along the polar b-axis and distinguishes γ-shear from other modes' contributions. The mixed-index $(\bar{1}32, 13\bar{2})$ pair contains contributions from multiple distortions and serves as a cross-check for the mode assignment. Together, these reflections form a minimal yet complete basis for identifying the observed oscillations.

Fig. 2 (a-c) shows time traces of relative intensity changes, $\Delta I(t)/I_0$, of the selected Friedel Bragg peak pairs $(\bar{4}08, 40\bar{8})$, $(0\bar{2}0, 020)$, and $(\bar{1}32, 13\bar{2})$, respectively. Here, $I_0$ denotes the averaged intensity before excitation, and $\Delta I(t) = I(t) - I_0$. All traces exhibit pronounced oscillations with different periods, which can be well described by a sum of cosine functions[31],

$$\frac{\Delta I(t)}{I_0} = A_0 e^{-(t-t_0)/\tau_0} + \sum_{i=1,2,3} A_i \cos[2\pi f_i(t - t_0) + \phi_i] e^{-(t-t_0)/\tau_i}, \qquad (1)$$

where $\tau_0$ represents the time constant of the lattice cooling, $t_0$ denotes the effective pump–probe time zero, at which the acoustic modes are launched, and $\tau_i$ ($i = 1, 2, 3$) are the decay times of the three modes, respectively. Cosine rather than sine basis functions are used because the extracted phases cluster near 0° or 180° [Fig. 2(d–e)], indicating that all three modes are launched displacively. In this regime, polarization suppression and lattice heating by photoexcited carriers produce a step-like stress[27, 31-33] that shifts the lattice equilibrium position on ultrafast timescales, such that the phonon coordinate begins at maximum displacement.

The fitted amplitudes and phases, as shown in Fig. 2 (d-f), allow unambiguous assignment of the three acoustic modes based on their symmetry-selective modulation of different Friedel Bragg pairs. For the $(\bar{4}08, 40\bar{8})$ pair [Fig. 2 (d)], the oscillations are dominated by the $f_1$ component, with a small contribution from the $f_3$ component, while the contributions from $f_2$ are negligible. Crucially, the $f_1$ oscillations of the two Friedel-related reflections are out of phase by nearly 180 degrees, indicating an antisymmetric lattice distortion that reverses sign upon inversion of the scattering vector. Such behavior is characteristic of a transverse shear distortion within the a-c plane. Thus, $f_1$ is assigned to a $\beta$-shearing mode, as illustrated schematically in Fig. 2 (g). In contrast, the $(0\bar{2}0)$ and $(020)$ pair [Fig. 2 (e)] exhibits oscillations at $f_2$, with the two reflections again oscillating out of phase, while the $f_1$ contribution is strongly suppressed. This selectivity reflects the sensitivity of (0k0) reflections to shear distortions involving relative displacements along the polar b-axis, and we therefore assign $f_2$ to a transverse $\gamma$-shearing mode [Fig. 2(h)]. Finally, the mixed-index $(\bar{1}32, 13\bar{2})$ pair [Fig. 2(f)] exhibits contributions from all three modes, with $f_1$ and $f_2$ remaining out of phase between the Friedel-related reflections, consistent with the assignments above. Notably, the $f_3$ component oscillates in phase for all examined Friedel pairs, consistent with a longitudinal distortion that preserves inversion symmetry of the scattering vector. We therefore identify $f_3$ as a longitudinal acoustic (LA-breathing) mode, as schematically illustrated in Fig. 2(i).

At first glance, the observation of an $f_3$ component (LA-breathing mode) in the measured $\Delta I(t)/I_0$ of the $(0\bar{2}0, 020)$ reflections appears counterintuitive, since the LA-breathing eigen-displacement for propagation along $a^*$ has a negligible projection along the $b$-axis, as confirmed by Christoffel-equation analysis (section S5). However, this discrepancy arises from experimental factors relevant to UED measurements on thin freestanding crystals. Owing to the finite transverse coherence, angular divergence, and energy spread of the pulsed electron beam, diffracted intensities extend over a finite excitation-error range in reciprocal space, producing a disk width of 0.7 nm$^{-1}$ for the $020$ reflection [Fig. 1(b)]. This finite excitation-error tolerance is further evidenced by static electron-diffraction measurements, which show that the same diffraction pattern remains observable when the sample is tilted by up to 8° away from the nominal [201]-zone-axis condition, although the intensities of individual reflections vary (section S6 and fig. S4). Given the angular separation of ~ 7° between the (000) → $(120)/(1\bar{2}0)$ directions and the [201] zone plane, together with curvature-induced local tilts of the freestanding flake, intensity modulations of nearby $(1\bar{2}0)/(120)$ reflections that are sensitive to the LA-breathing distortion can partially contribute to the $(0\bar{2}0)/(020)$ integration window (section S6 and fig. S4). This overlap gives rise to an apparent in-phase $f_3$ contribution in the measured $\Delta I(t)/I_0$, as illustrated in Fig. 2(i), thereby reconciling the observation of the LA-breathing mode in $(0\bar{2}0, 020)$ reflections without contradicting its intrinsic displacement symmetry.

The above mode assignment is further supported by comparing their corresponding acoustic velocities. Based on the measured mode frequencies $f_1 = 8 \pm 0.3$ GHz, $f_2 = 9.3 \pm 0.3$ GHz, $f_3 = 12.7 \pm 0.3$ GHz [Fig. 1 (d-g)], and the flake thickness $d$ = 100 nm, we can calculate the corresponding acoustic wave velocities $v = 2df$. This yields $v_\beta$ = 1.6 $\pm$ 0.1 km/s, $v_\gamma$ = 1.9 $\pm$ 0.1 km/s, and $v_{LA}$ = 2.5 $\pm$ 0.1 km/s. The LA-breathing mode propagates faster than the two shear modes, consistent with the elastic stiffness of the monoclinic lattice[34]. Additionally, using the stiffness tensor obtained from density functional theory (DFT) calculations[34], we computed the ratio of corresponding acoustic velocities (section S5) and found with the ratio $v_{LA}/v_\beta \approx 1.55$ agrees well with experimental value of 1.56.

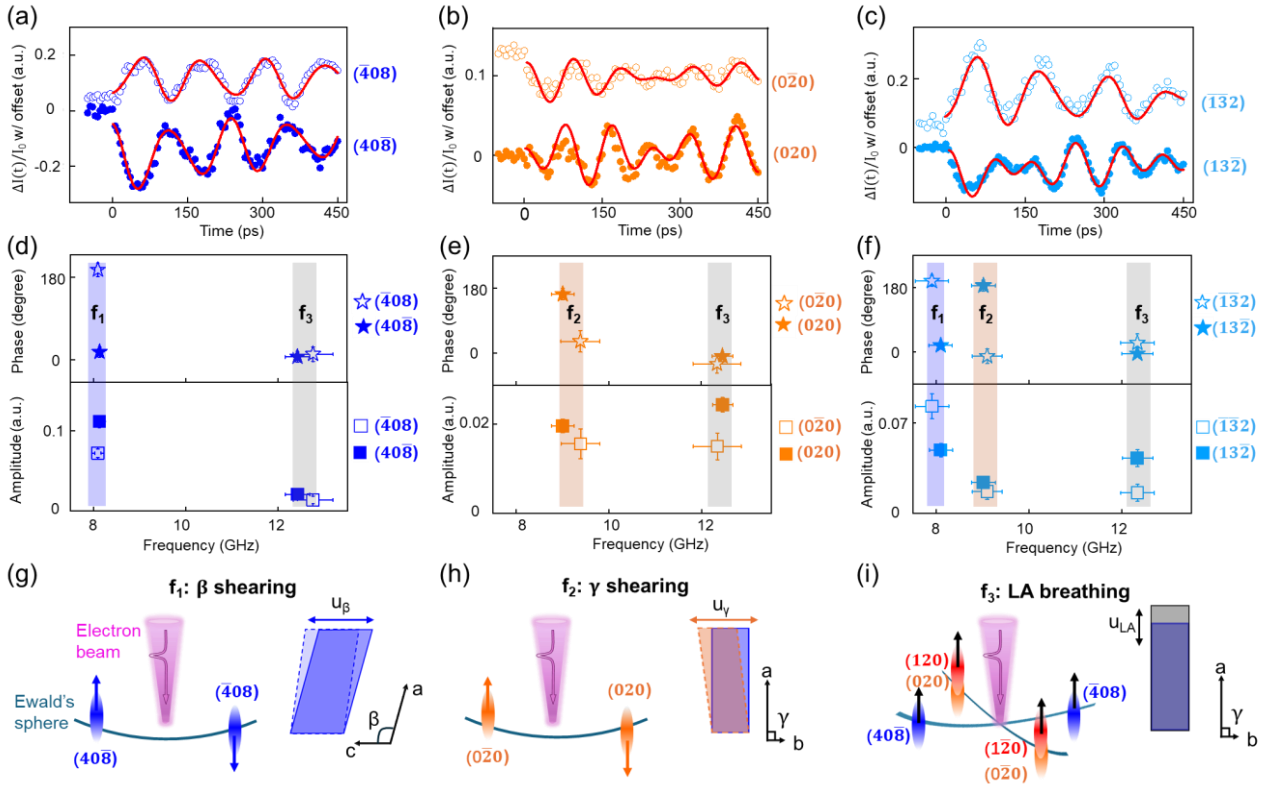

*Fig. 2. Reciprocal-space identification of three coherent acoustic modes in NbOI₂. (a, b, c) Time-resolved diffraction intensity changes, ΔI/I₀, of three representative Friedel Bragg peak pairs ($\bar{4}08, 40\bar{8}$; $0\bar{2}0, 020$; $\bar{1}\bar{3}2, 13\bar{2}$) measured on the 100-nm-thick flake. Red curves are multi-cosine fits according to Eq. (1). (b, e, f) Extracted amplitudes and phases for the three frequency components. f₁*

*and $f_2$ exhibit opposite phases between Friedel Bragg pairs, whereas $f_3$ remains approximately in phase across the intensity changes of all Bragg peaks. Friedel pair ($\bar{4}08, 40\bar{8}$) is dominated by $f_1$ with a small component of $f_3$; ($0\bar{2}0, 020$) are dominated by $f_2$ and $f_3$, with negligible $f_1$; ($\bar{1}32, 13\bar{2}$) show mixed multi-frequency responses. (g-i). Schematic illustrations of β-shearing, γ-shearing, and LA-breathing distortions and their modulation of the corresponding Friedel pairs, establishing the mode assignments. Panel (i) highlights that the apparent $f_3$ contribution in ($0\bar{2}0, 020$) can arise from intensity mixing to nearby Bragg peaks ($1\bar{2}0, 120$).*

**Spatial heterogeneity in acoustic mode amplitudes and lifetimes**

To visualize the spatial distribution of the acoustic modes, we computed FFT amplitude maps from time-resolved UEM images using 90 nm × 90 nm analysis windows, substantially smaller than the micrometer-scale ferroelectric domain size of NbOI2[22, 27]. This enables resolving nanoscale variations in mode content that would be obscured by spatial averaging across domain structures, where domain walls can scatter acoustic phonons and modify elastic responses[35-37]. Figures 3(a–c) display the FFT amplitude maps at $f_\beta$ = 8 GHz (β-shearing), $f_\gamma$ = 9.3 GHz (γ-shearing), and $f_{LA}$ = 12.7 GHz (LA-breathing), respectively. Three representative Regions (I–III), outlined by light-blue, yellow, and white dashed contours, exhibit pronounced spatial heterogeneity of the acoustic response. To quantify the spatial distribution of the three modes, we define the normalized mode amplitude $\delta_i$ in each Region as the integrated FFT amplitude of that mode within the Region, normalized by its total integrated amplitude over the entire frame. As shown in Fig. 3(d), the β-shearing response is strongly concentrated in Region I, where contributions from the other two modes are negligible; it exhibits a reduced presence in Region II and is nearly absent in Region III. In contrast, Regions II and III display progressively larger FFT amplitudes at $f_\gamma$ and $f_{LA}$, showing an increased relative weight of γ-shearing and LA-breathing dynamics. This spatial heterogeneity reflects variations in how the optical excitation couples to different acoustic eigenmodes at the nanoscale, likely influenced by local mechanical conditions, domain structures, and structural inhomogeneities, leading to a non-uniform distribution of excitation among transverse and longitudinal modes across the freestanding flake.

To examine how this heterogeneity influences energy dissipation, we analyzed $\Delta I(t)/I_0$ traces from 33 smaller regions of interest (ROIs) across Regions I–III (section S7 and fig. S5). Fig. 3(e) presents three representative time traces—one from each Region—marked by yellow rectangles in Fig. 3(a). The oscillation in ROI$_1$, taken from Region I, is dominated by the β-shearing mode and exhibits a noticeably longer decay time than those in ROI$_2$ and ROI$_3$. In contrast, ROI$_2$ contains comparable contributions from all three modes, while ROI$_3$ is primarily governed by γ-shearing and LA-breathing dynamics, as confirmed by their FFT spectra [Fig.

3(f)]. All time traces are well described by damped cosine functions [Eq. (1)], enabling extraction of mode-specific decay times (section S7 and fig. S5). Fig. 3(g) displays the statistically averaged decay times extracted from a total of 33 ROIs across Regions I–III. The β-shearing mode exhibits the longest mean decay time in Region I, $\overline{\tau_\beta}$ = 1.35 ± 0.15 ns, which is reduced to 0.69 ± 0.09 ns in Region II, where multimode coexistence is strongest. Here, the error bars denote the standard error of the mean. This reduction is consistent with enhanced phonon–phonon scattering and decoherence in regions with mixed mode populations. In Regions II and III, the LA-breathing mode, when present, consistently shows longer decay times than the shear modes, indicating greater robustness of longitudinal motion. In Region III, closest to the supporting Si grid, both LA-breathing and γ-shearing modes are further reduced compared to Region II, consistent with efficient substrate-mediated energy leakage. Our results suggest that acoustic phonon–phonon scattering and substrate coupling provide dominant dissipation pathways that limit acoustic coherence in freestanding $NbOI_2$.

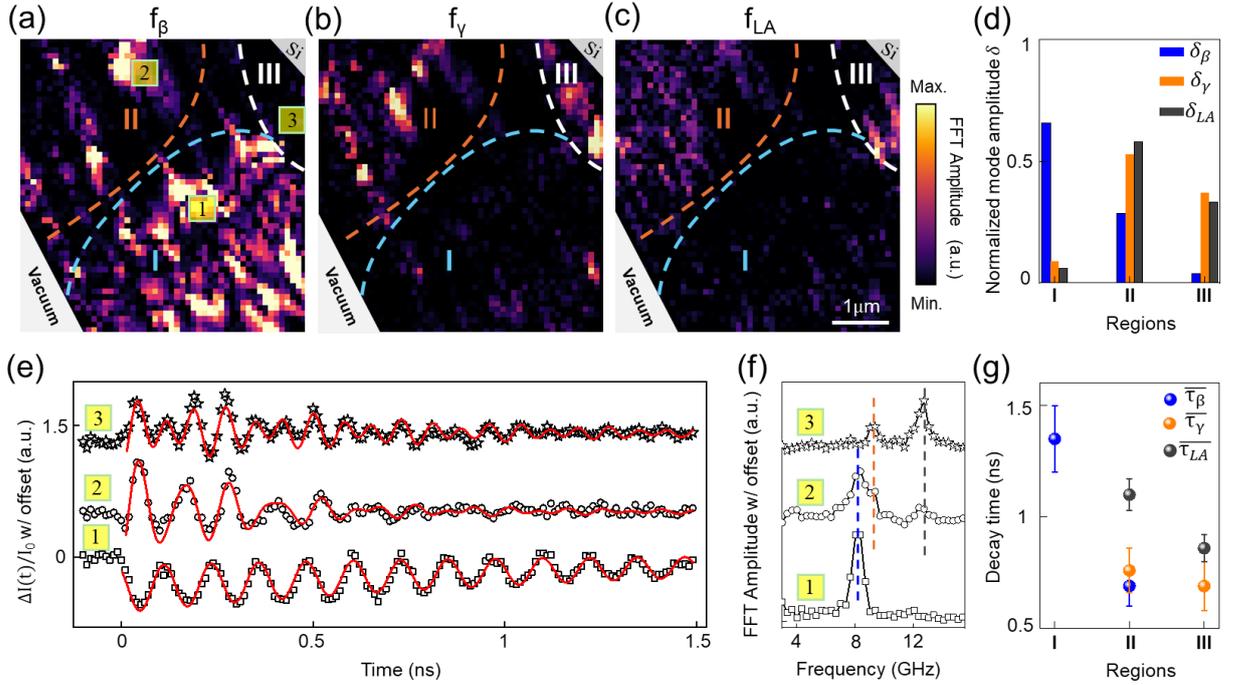

*Fig. 3. Spatially resolved coherent acoustic modes and their decay dynamics.* (a-c) FFT amplitude maps at $f_\beta$ = 8 GHz, $f_\gamma$ = 9.3 GHz, and $f_{LA}$ = 12.7 GHz, revealing pronounced spatial heterogeneity. Regions I–III (dashed contours as guides of the eye) exhibit distinct mode distributions: Region I is dominated by β-shearing, Region II shows the coexistence of all three modes, and Region III is dominated by the γ-shearing and LA-breathing modes. Despite less sensitivity to LA motion along the [201] zone axis, its relative enhancement in Regions II and III is resolved. (d) Normalized mode amplitude in Regions I–III. (e). Representative $\Delta I(t)/I_0$ traces from three ROIs (yellow boxes labeled 1–3 in panel a). Red curves are fitted results. (f). FFT spectra of the traces in (e), with dashed lines

*marking f<sub>β-shear</sub>, f<sub>γ-shear</sub>, and f<sub>LA-breathing</sub>, respectively. (g) Area-averaged decay times of the three modes in Regions I–III, showing longer β-shearing lifetimes in Region I and reduced shear-mode lifetimes in Regions II and III, where LA-breathing mode is more prominent.*

**Discussion**

To elucidate the physical origin of the three observed acoustic modes, we distinguish the roles of transient depolarization and lattice heating in driving the lattice response. Given the comparable detection sensitivity to the two TA modes under the [201] zone-axis geometry, the pronounced dominance of the β-shearing response over the γ-shearing [Fig. 3 (a, b)] reflects a highly anisotropic interaction between the ultrafast depolarization and shear strain. Pure lattice heating can be excluded as the primary driver of the TA modes: the laser spot (~ 50 μm) is much larger than the freestanding window (~ 5 μm), yielding a laterally uniform temperature rise that suppresses in-plane thermal gradients and associated thermoelastic shear stresses. We therefore attribute the generation of the TA modes to the transient depolarization field, which launches shear stress via the inverse piezoelectric effect[22,34,38]. In monoclinic NbOI$_2$, the relevant piezoelectric tensor components couple the depolarization field $\Delta E$ to shear strains as $\epsilon_{ac} \propto e_{25}\Delta E$ (β-shearing) and $\epsilon_{ab} \propto e_{26}\Delta E$ (γ-shearing), with $\Delta E \sim 7 \times 10^8$ V m$^{-1}$. Previous theoretical studies consistently report larger $e_{25}$ than $e_{26}$. For example, Ref. [34] reports a bulk value of $e_{25} = 5\times 10^{-3}$ C/m$^2$ and $e_{26} \sim 0$ C/m$^2$. Another theoretical study[22] reports a sheet value of $e_{26} = 0.7\times 10^{-10}$ C/m, which corresponds to 0.7 $\times 10^{-3}$ C/m$^2$ for our 100 nm-thick flake, and remains substantially smaller than the reported $e_{25}$. This strong anisotropy in the inverse piezoelectric coupling explains the observed dominance of the β-shearing mode.

In contrast, piezoelectric coupling between the depolarization field $\Delta E$ (along the polar b-axis) and the out-of-plane normal strain $\epsilon_{zz}$ is symmetry-forbidden or negligibly small in the monoclinic C2 structure, suggesting that the transient depolarization selectively drives the shear modes but not the LA-breathing mode. Although quadratic electrostrictive coupling may also contribute, it is generally weaker than the linear piezoelectric response[39]. The LA-breathing mode instead originates primarily from thermoelastic stress generated during carrier thermalization and energy transfer to the lattice, producing an out-of-plane pressure $\sigma_{thermal} \propto C_{33}\alpha_c \Delta T$ that drives the breathing oscillation[40,41], where $C_{33}$ denotes the out-of-plane longitudinal elastic stiffness, and $\alpha_c$ is the linear thermal expansion coefficient along the surface normal. The relatively weak LA amplitude in both diffraction and UEM data reflects the reduced detection sensitivity to normal displacements under the [201] zone-axis geometry. The coexistence of depolarization-driven shear motion and thermally driven longitudinal motion demonstrates that distinct nonequilibrium pathways launch different acoustic symmetries on comparable timescales. More broadly, this separation of driving mechanisms provides a microscopic basis for mode-selective elastic excitation in low-dimensional ferroelectrics.

Beyond the mode-selected generation pathways, the real-space measurements reveal the nanoscale spatial heterogeneity of the launched modes. The spatial correlation between local mode composition, amplitude, and decay shows that acoustic energy dissipation is not intrinsic to a single phonon branch but is determined by the coexistence of different acoustic symmetries within a nanoscale region, where phonon–phonon scattering provides an efficient dissipation channel. Our results establish a microscopic, mode-resolved picture of ultrafast electromechanical dynamics and elastic energy dissipation in $NbOI_2$, providing a framework for understanding and tuning acoustic coherence in low-dimensional polar materials and devices.

## MATERIALS AND METHODS

### Sample preparation

$NbOI_2$ flakes were mechanically exfoliated from a bulk crystal using scotch tape and dry-transferred onto a SiN grid (NT0005, Norcada). The bulk $NbOI_2$ crystal was synthesized by chemical vapor transport[25]. Flake thickness was determined by atomic force microscopy. Optical and Raman characterizations are provided in section S1.

### Ultrafast electron microscopy and diffraction

The UEM and UED experiments were carried out at the Center for Nanoscale Materials, Argonne National Laboratory[32]. The experiments employed a stroboscopic pump–probe configuration with femtosecond optical excitation and synchronized electron pulses. The pump pulse (515 nm, 400 fs) operated at a repetition rate of 50 kHz with a spot size of ~50 μm. The probe consisted of 200-keV electron pulses, providing ~1 ps temporal resolution and ~1 nm real-space spatial resolution for imaging, together with selected-area electron diffraction for reciprocal-space measurements. All data shown in the main text were acquired at room temperature under an absorbed fluence of ~3 $mJ/cm^2$. Real-space UEM imaging and diffraction were performed on the same freestanding flake by switching between imaging and diffraction modes of the microscope, enabling direct correlation between nanoscale lattice motion and symmetry-resolved structural dynamics.

## Acknowledgement

**Funding**: This work was primarily supported by the U.S. Department of Energy, Office of Science, Basic Energy Sciences, Materials Sciences and Engineering Division, under Awards no. DE-SC0012509. Y. L. acknowledges the support by the U.S. Department of Energy, Office of Science, Basic Energy Sciences, Materials Sciences and Engineering Division, under contract no. DEAC02-06CH11357. Work performed at the Center for Nanoscale Materials, a U.S. DOE Office of Science User Facility, was supported under Contract DEAC02-06CH11357. C.F. and J.X. acknowledge support from NSF-DMR-2237761. Z.Z., P.Y., and B.L. acknowledge support from NSF-25163643, ONR Grant N00014-23-1-2020, and AFOSR FA9550-19-1-0037.

**Author contributions**: H. W. and J. X. conceived the project. Z. C., T. G., H. L., and Y. L. performed the UEM and UED measurements. Z. Z., P. Y., and B. L. grew the NbOI$_2$ bulk crystals. C.F. prepared the thin NbOI$_2$ flakes on Si grids under the guidance of J. X.. Z. C. analyzed the data and wrote the manuscript with input from all authors

**Competing interests:** The authors declare that they have no competing interests.

**Data and materials availability:** All data needed to evaluate the conclusions in the paper are present in the paper and/or the Supplementary Materials.

# Supplementary Materials

**S1: Optical, thickness, and Raman characterization of the NbOI$_2$ flake.**

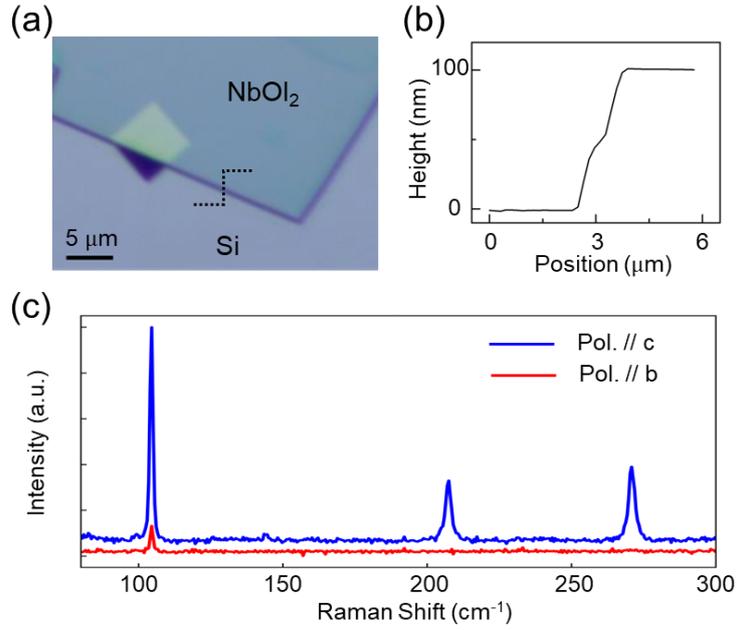

*Fig. S1. (a) Optical image of an NbOI$_2$ flake suspended on a Si grid. (b) AFM height profile measured along the dashed line in (a), identifying the sample thickness of 100 nm. (c) Raman spectra of the NbOI$_2$ flake with the incident linear polarization parallel to the b-axis (red) and c-axis (blue). The excitation wavelength is 532nm, and the power is 1 mW, with a ~ 0.5 um spot size on the sample. The sample is rotated to align the b-axis or c-axis parallel to the polarization of the laser. The intensity of the $A_g$-like modes shown below (~ 104 cm$^{-1}$, ~ 208cm$^{-1}$, ~ 270cm$^{-1}$) is quenched when the incident polarization is along the b-axis and maximized when along the c-axis, consistent with the previous report [S1].*

## S2: Ultrafast suppression of polarization.

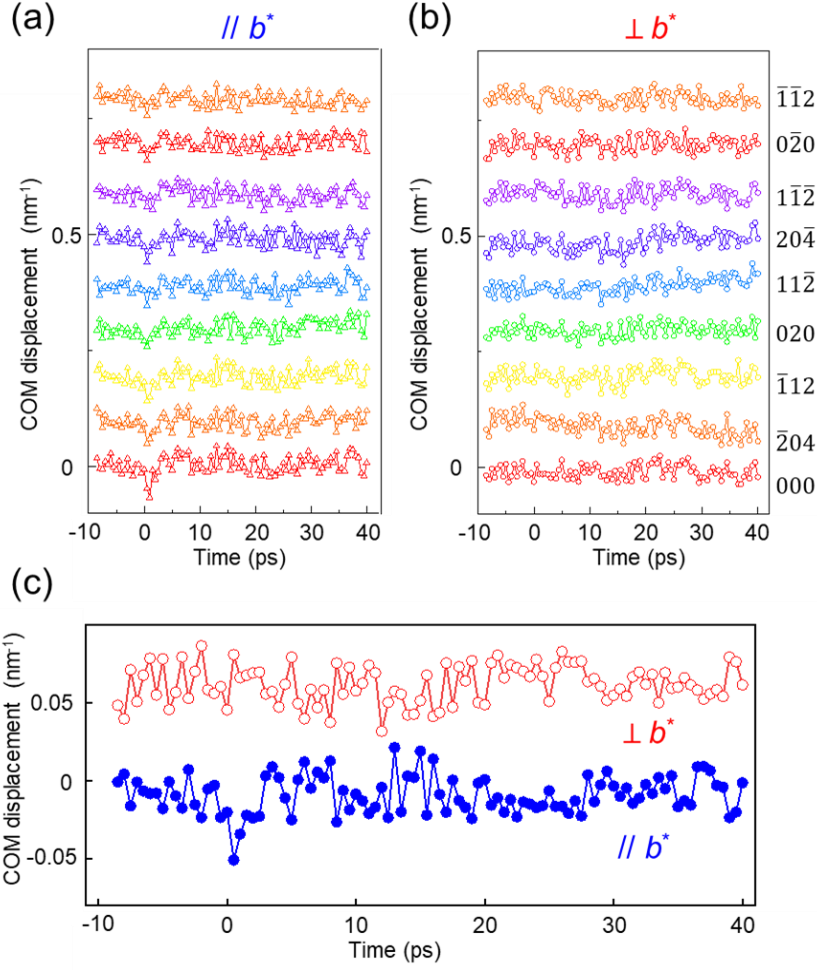

*Fig. S2. Laser excitation induced center-of-mass (COM) displacement of Bragg peaks in NbOI2. (a) Time-dependent COM displacement of nine representative Bragg peaks projected along the **b\*** direction, showing a pronounced negative shift immediately following photoexcitation. (b) COM displacement of the same Bragg peaks projected perpendicular to **b\***, where no systematic transient shift is observed. (c) Average COM displacement of all nine Bragg peaks along **b\*** (blue) and perpendicular to **b\*** (red). The averaged data highlight a maximum transient shift of ~0.05 nm$^{-1}$ toward −**b\***, indicating a pump-induced rigid translation of the diffraction pattern consistent with transient electron-beam deflection by a lateral electric field associated with photoinduced polarization suppression.*

A quantitative estimate of the maximum transient change in the electric field ***E*** associated with photoinduced polarization suppression is obtained from the pump-induced rigid shift of the electron diffraction pattern. As shown in Fig. S2a, the center-of-mass (COM) positions of nine Bragg peaks aligned parallel to the ***b\**** direction exhibit a pronounced negative shift immediately following excitation, whereas no systematic shift is observed for projections perpendicular to ***b\**** (Fig. S1b). The average COM displacement of all nine Bragg peaks (Fig. S1c) reveals a maximum transient shift of $\Delta k_{max} = 0.05$ nm$^{-1}$ toward -***b\****, confirming that the effect arises from a shift of the whole diffraction pattern rather than a change in lattice spacing. The corresponding electron-beam deflection angle is obtained from $\Delta k = k_0 \gamma$, where $k_0 = 2\pi/\lambda$ is the incident electron wavevector and $\lambda$ is the relativistic electron wavelength. For 200 keV electrons ($\lambda$ = 2.51 pm), this yields $k_0$ = 2.5 × 10$^3$ nm$^{-1}$ and a maximum deflection angle of $\gamma_{max}$ = 0.02 mrad. Using the relativistic relation $\gamma = em^*LE/p^2$, where $m^*$ and $p$ are the relativistic mass and momentum of the electrons, respectively, and $L$ = 10 nm represents the effective excitation (absorption) depth [S2], this deflection corresponds to a maximum transient change in the lateral electric field of $E_{max}$ ~ 7 × 10$^8$ Vm$^{-1}$. This field change is attributed to transient screening of the long-range electric field associated with the in-plane ferroelectric polarization along the *b* axis, providing a quantitative upper bound on the photoinduced polarization-suppression signal observed in the UED measurements.

### S3: A real-time video of the contour oscillations following laser excitation (Movie S1).

The attached UEM video shows bend-contour oscillations following laser excitation. The video displays frames that have been spatially drift-corrected and intensity-normalized. Spatial drift was corrected using stationary features present in all UEM images, and the intensity of a vacuum region was used as a reference for normalization. All UEM data presented in the main text are shown after drift correction and normalization.

### S4: Laser polarization-dependent contour response.

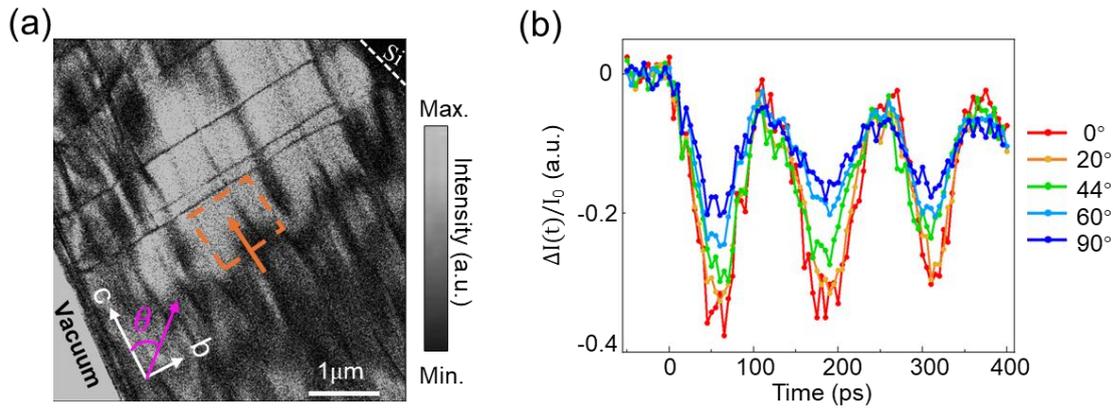

*Fig. S3. (a) Bright-field UEM image acquired before laser excitation. The purple arrow indicates the polarization direction of the 515-nm pump laser. The angle θ is defined as the angle between the laser polarization and the crystallographic c-axis. The orange dashed rectangle marks the region where the bend contour oscillates along the direction indicated by the orange arrow after excitation. (b) Polarization-angle-dependent relative intensity change, ΔI/I₀, extracted from the region marked by the dashed rectangle in (a). The ΔI/I₀ amplitude is maximized when the laser polarization is aligned with*

*the c-axis (θ = 0°), and is approximately a factor of two larger than that obtained when the polarization is along the b-axis (θ = 90°), consistent with the strongly anisotropic optical absorption of NbOI₂ [S1].*

**S5: Polarization of Acoustic Eigenmodes from the Christoffel Equation.**

To clarify the polarization character of the LA-breathing mode discussed in the main text, we analyze the acoustic eigenmodes of monoclinic NbOI$_2$ using the Christoffel equation formalism [S3]. This approach determines the propagation velocities and displacement polarizations of elastic waves in anisotropic crystals directly from the elastic stiffness tensor. For a plane acoustic wave propagating along a direction $\bm{n}$, the Christoffel equation is given by $\sum_{j,l} \Gamma_{ij}(\bm{n}) u_j = \rho v^2 u_i$ and $\Gamma_{ij}(\bm{n}) = \sum_{k,l} C_{ikjl} n_k n_l$, where $C_{ikjl}$ is the elastic stiffness tensor, $\rho$ is the mass density, $v$ is the velocity, and $\bm{u}$ is the displacement eigenvector of the acoustic mode. We quote the stiffness matrix from the density functional theory work of NbOI$_2$ [S3]:

$$C = \begin{pmatrix} 11.3 & 7.91 & 7.13 & 0 & -0.07 & 0 \\ 7.91 & 330.54 & 6.19 & 0 & 0.56 & 0 \\ 7.13 & 6.19 & 89.07 & 0 & -10.12 & 0 \\ 0 & 0 & 0 & 18.64 & 0 & -1.37 \\ -0.07 & 0.56 & -10.12 & 0 & 6.12 & 0 \\ 0 & 0 & 0 & -1.37 & 0 & 4.33 \end{pmatrix},$$ where the units of each term in the matrix are all in GPa.

In an orthonormal basis ($\hat{x}$, $\hat{y}$, $\hat{z}$) chosen such that $\hat{y}$ || b-axis and $\hat{x}$-$\hat{z}$ spans the a-c plane, propagation along $\bm{a^*}$ has the form $\bm{n} = [\cos(15.5°), 0, \sin(15.5°)]$, which is perpendicular to b-axis. Using the NbOI$_2$ elastic constants, we obtain (in GPa)

$$\Gamma(\bm{n}||\bm{a^*}) = \begin{pmatrix} 10.58 & 0 & 1.84 \\ 0 & 1.16 & 0 \\ 1.84 & 0 & 5.18 \end{pmatrix}.$$ Diagonalizing $\Gamma$ yields three eigenvectors $\bm{u}$: one transverse mode with displacement along the b-axis, $\bm{u}_\gamma = (0, 1, 0)$, corresponding to the γ-shearing; one transverse mode (i.e., β-shearing) with $\bm{u}_\beta = (-0.294, 0, 0.956)$ and one LA-breathing mode with $\bm{u}_{LA} = (-0.956, 0, -0.294)$. Both $\bm{u}_\beta$ and $\bm{u}_{LA}$ have no displacement component along the b-axis.

The relative wave velocities of the three acoustic branches propagating along $\bm{a^*}$ are obtained from eigenvalues $\lambda_i$ of the Christoffel matrix via $v_i = \sqrt{\lambda_i/\rho}$. As the velocity ratios depend only on elastic anisotropy and are independent of the mass density, $v_i/v_j = \sqrt{\lambda_i/\lambda_j}$.

By solving the eigenvalues of the above Christoffel matrix, the calculated velocity ratio between LA-breathing and $\beta$-shear is 1.55, perfectly matching the experimental one (1.56).

## S6: Diffraction pattern as a function of incident angle.

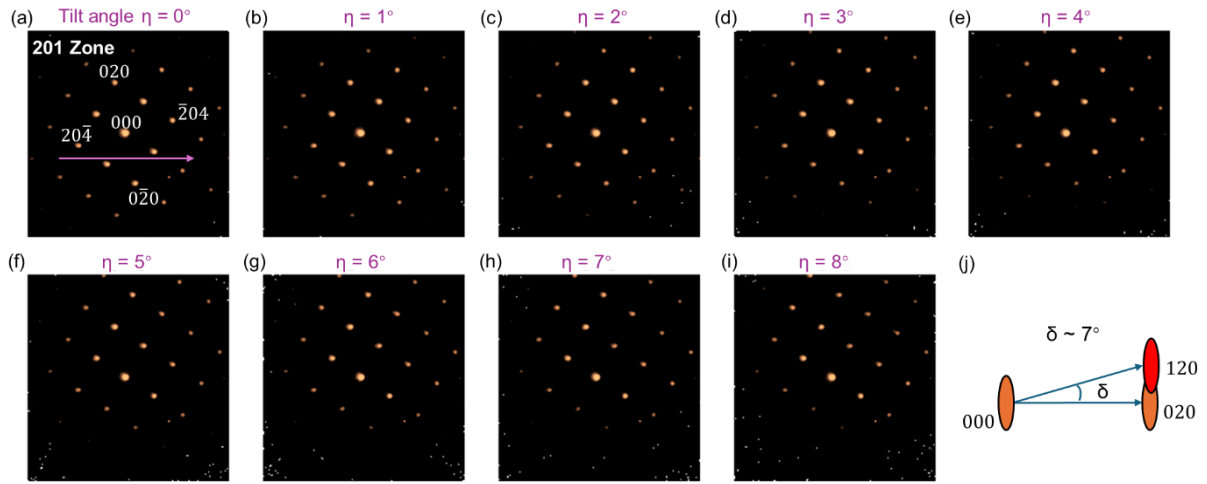

*Fig. S4. Static electron diffraction measurements demonstrating the angular tolerance of Bragg reflections under the nominal [201] zone-axis condition. (a) Electron diffraction pattern acquired at the nominal [201] zone axis. (b-i) Diffraction patterns acquired after rotating the sample incrementally by 1° to 8° about the axis indicated by the purple arrow in (a). Over this angular range, the diffraction pattern remains unchanged, and the same set of Bragg reflections remains detectable, while the intensities of individual reflections vary with tilt angle. (j) Schematic illustration of the reciprocal-space geometry, showing the angular separation between the (000)→(020) and (000)→(120) directions to be approximately 7°. The persistence of Bragg patterns over comparable angular deviations supports the partial contribution of the neighboring (120) reflection to the (020) integration window in the UED measurements.*

# S7: Complete set of time-resolved $\Delta I/I_0$ traces for all ROIs

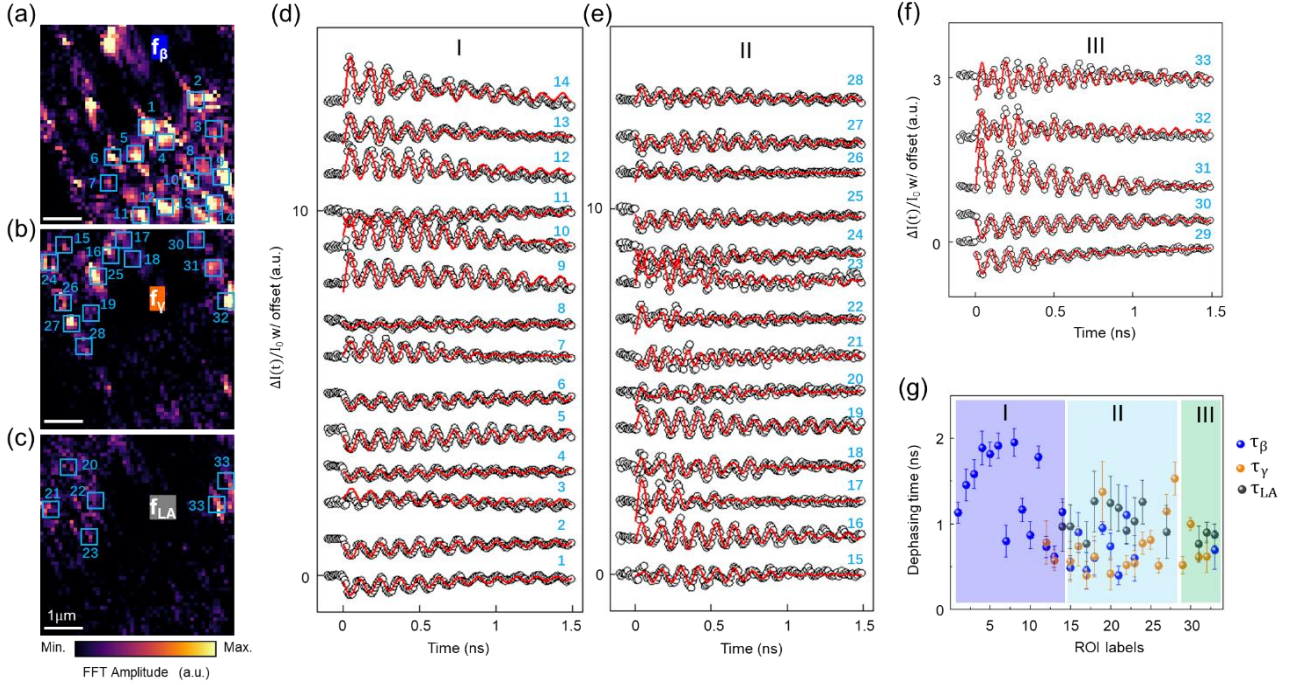

*Fig. S5. Full set of time-resolved $\Delta I/I_0$ traces and decay statistics for all ROIs. (a–c) FFT amplitude maps at the three characteristic frequencies ($f_\beta$, $f_\gamma$, and $f_{LA}$), reproduced from Fig. 3, with all analyzed regions of interest (ROIs) marked. ROIs 1–14 are located in Region I, ROIs 15–28 in Region II, and ROIs 29–33 in Region III. (d–f) Complete set of time-resolved $\Delta I(t)/I_0$ traces (open symbols) and corresponding fits (red curves) for all ROIs in Regions I, II, and III, respectively, using Eq. (1). (g) Extracted decay times for all ROIs and modes, grouped by Region, showing the statistical distribution underlying the averaged values presented in Fig. 3(f).*